\documentstyle[12pt,graphicx]{article}
\begin{document}

\title{What can the quantum liquid say on
the brane black hole, the entropy of extremal black hole and
the vacuum energy?}
\author{G.E. Volovik\\
~~\\
  Low Temperature Laboratory,
Helsinki University of Technology\\
   P.O.Box 2200, FIN-02015
   HUT, Finland \\
and\\
  Landau Institute for Theoretical Physics\\
  Kosygina 2,
117334 Moscow, Russia
}

\maketitle
Using quantum liquids one can simulate the behavior of the
quantum vacuum in the presence of the event horizon. The
condensed matter analogs demonstrate that in most cases the
quantum vacuum resists to formation of the horizon, and even if
the horizon is formed different types of the vacuum instability
develop, which are faster than the process of Hawking
radiation. Nevertheless, it is possible to create the horizon on
the quantum-liquid analog of the brane, where the vacuum
life-time is long enough to consider the horizon as the
quasistationary object. Using this analogy we calculate the
Bekenstein entropy of the nearly extremal and extremal black
holes, which comes from the fermionic microstates in the region
of the horizon -- the fermion zero modes. We also discuss how
the cancellation of the large cosmological constant follows from
the thermodynamics of the vacuum.

\section{Introduction}

In general relativity the event horizon is a rather fragile
construction:  small deviations from Einstein theory leads to the
catastrophical consequence for the event horizon.  In
particular,  the horizon is destroyed: (i) after inclusion of
arbitrarily small mass terms for graviton
\cite{BabakGrishchuk};   (ii) after
inclusion of higher order curvature terms \cite{Holdom}; (iii) in the
models allowing varying speed of light \cite{Magueijo}; etc.
Thus there is an open question whether the horizon can be really
formed.

The condensed matter analogs of gravity also indicate that the
event horizon for quasiparticles is a subtle issue: in most
systems it cannot be formed. The characteristic example is the
  black-hole horizon for sound waves generated by a moving
liquid as suggested by Unruh
\cite{Unruh}.  The horizon for sound
waves can be formed when the local velocity of the liquid
exceeds the speed of sound. In liquids, however, the effective
metric experienced by sound waves obeys the hydrodynamic
equations which thus play the role of Einstein equations,
and these hydrodynamic equations  prohibit the existence of the
spherically symmetric event horizon
\cite{VolovikReview}. The horizon can be formed only in the
vessels of special geometry such as the so-called Laval nozzle
\cite{LavalNozzle2002}.  However, even in this case the horizon
will be spoiled by the formation of shock waves.

The constraints dictated by the hydrodynamic equations can be
avoided if one constructs the horizon for such quasiparticles
whose maximum attainable speed $c$, which plays the role of the
`speed of light', is smaller than the speed of sound. In this
case the speed of sound is not reached at the quasiparticle
horizon, and thus the horizon is not accompanied by the
hydrodynamic instability. Such situation can occur, for example,
in superfluids $^3$He-A where the superfluid ground state plays
the role of the quantum vacuum, while some low-frequency
bosonic collective modes of the superfluid vacuum mimic the
effective gravitational field experienced by the `relativistic'
quasiparticles -- the Weyl fermions
\cite{VolovikBook}. In this superfluid liquid the effective
gravity is very similar to the Sakharov gravity
\cite{Sakharov} induced by quantum fluctuations of the vacuum.

In $^3$He-A the `speed of light'
$c$  is about three order of magnitude smaller than the speed of
sound.   Nevertheless the other instability becomes important --
the instability of the superfluid quantum vacuum when it moves
with supercritical velocity, $v>c$.
This brings us to the problem of the stability of the quantum
vacuum in the presence of the non-trivial metric field.
After the first proposal by Unruh there were many projects in
condensed matter and optics to simulate the event horizon
(see the book \cite{ArtificialBlackHole}). Till now we
were not able to construct, even in the gedanken experiment, the
condensed matter system whose vacuum state remains stable in the
presence of the horizon. Who knows, maybe this is an indication
that the real quantum vacuum -- the ether -- is also highly
sensitive to the presence of the horizon or ergoregion.

However,
recently it was found that there exists a real and experimentally
accessible system where the vacuum instability will be developed
rather slowly, so that we can investigate the behavior of the
vacuum in the presence of the horizon. This is the horizon
constructed for quasiparticles living in the analog of the brane
-- the 2+1 world of the interface between two bulk superfluid
vacua, superfluid $^3$He-A and superfluid
$^3$He-B, which we refer to as the AB-brane.
In this situation the hydrodynamic equations determine the flow
of two superfluids in bulk, and the  velocities of this
flow near the AB-brane dictate the effective metric for
quasiparticles living on the brane -- the ripplons --   which
play the role of the brane matter. The maximum attainable
velocity
$c$ for the ripplons is typically much smaller than any
critical velocity  for the instability of the bulk superfluids.
As a result, the  ergoregion and the event horizon can be easily
constructed without any instability in the bulk liquids.

While the bulk vacuum remains stable, the brane vacuum is
nevertheless unstable beyond the horizon. This instability
occurs due to the interaction  of quasiparticles living on
the brane with that living in the bulk liquid (i.e. in the
higher-dimensional space outside the brane) which leads to the
decay of the brane vacuum in the region behind the horizon. The
weaker is the interaction the longer the black hole lives. This
mechanism of the vacuum decay can be applicable for the
astronomical black holes, if we really live on brane of the
higher-dimensional world. If the matter fields on the brane are
properly coupled to, say, gravitons in the bulk, this may lead
to the collapse of the black hole which is  fast as compared to
the popular mechanism of the black-hole evaporation by the
Hawking radiation \cite{HawkingNature}. This is another example
of the fragility of the horizon, now due to the interaction with
the higher-dimensional world. We shall consider this example in
Section 2.

If the time of the collapse of the black-hole horizon is long
enough, we can study the thermodynamics related to the horizon:
its temperature and the Bekenstein entropy.  In condensed matter
systems, as well as in quantum vacuum, the low-temperature
thermodynamic properties of the system is determined by  zero
modes, fermionic or/and bosonic. If the condensed matter
system has no zero modes, the gapless (massless)
modes necessarily appear in the presence of the ergoregion or
event horizon. In such systems, all the entropy comes from the
microstates in the black hole interior or in the vocinity of the
horizon, where the fermion zero modes provide the source of the
Bekenstein entropy \cite{Bekenstein}. In Section 3 we consider
here the simplest example of the entropy produced by the
condensed-matter analog of the extremal black hole.

The condensed matter examples demonstrate that the effective
gravity can essentially differ from the fundamental gravity even
in principles. For example, in the effective gravity the
effective spacetime can be geodesically incomplete. This is
simply because the description of the quasiparticle motion in
terms of the metric field is incomplete.  The effective gravity
appears only in the low-energy corner:  the general covariance
and Einstein equations even if they were valid at low energy,
are violated at high energy. The quasiparticle thus can simply
leave the low-energy world and enter the non-symmetric
high-energy world where the quasiparticle trajectories are
governed by the non-relativistic and non-covariant laws.

Since in the effective gravity the general covariance is lost at
high energy, the metrics which for the low-energy observers look
as equivalent, since they can be transformed to each other by
coordinate transformation, are not equivalent physically.
Moreover, the vacuum can be essentially different for different
metrics which are formally equivalent. If our gravity is also
effective, we can in principle resolve between the
formally equivalent metrics if we will be able to reach the high
enough energy.  For example, the Painlev\'e-Gullstrand
\cite{Painleve}, Schwarzschild and other metrics used for
description of the black hole could be discriminated on physical
ground. It is interesting that in the moving superfluids the
effective metric which describes the event horizon is of the
Painlev\'e-Gullstrand type where the time reversal symmetry is
violated. That is why we shall also use such a type of the
metric field for calculations of the Bekenstein
entropy  related to a
horizon. The event horizon which emerge in stationary
superfluids and which simulates the Schwarzschild black hole is
discussed in \cite{ChaplineLaughlin}.

We will be interested here mainly  in the entropy of the
extremal black hole, in particular because there is a
contradiction between different approaches to the
calculations of the entropy of such a hole. While the Bekenstein
entropy of the Reissner-Nordstr\"om black hole
in the limit of the extreme hole is proportional to the area of
the horizon,
${\cal S}=\pi r_+^2 p_0^2/\hbar^2$, where $p_0$ is the Planck
momentum, the entropy of the exactly extremal black hole is zero
according to
\cite{GibbonsKallosh,HawkingHorowitz}. Such a huge jump under
any small change of the extremal hole towards the non-extremal
one is rather problematic (see discussion in \cite{Page}).
We consider this problem using the effective gravity and the
natural regularization which comes from the
non-relativistic Planckian physics of quantum liquids at
high energy. We find that the Planckian physics modifies the
black hole entropy and smoothens the discontinuity occuring in
the limit of the extremal black hole. It happens that for the
calculations of the entropy of the extremal or nearly extremal
black holes it is not necessary to know the quasiparticle
spectrum in the trans-Planckian region: the regularization
occurs already at low energy where it is enough to use only the
first (cubic)  correction to the linear relativistic spectrum:
$E(p)\approx cp(1 +  p^2/p_0^2+\ldots)$.

The thermodynamic approach is also useful for discussion
of the condensed matter analogs of the cosmological constant
problems (the detailed discussion can be found in the book
\cite{VolovikBook}). In quantum liquids
these problems are essentially the same as in our quantum
vacuum, with probably the only difference that in the
quantum liquids the structure of the quantum vacuum is well
known even in the trans-Planckian region. In Section 4 we discuss
how the cosmological constant problems (why it is not huge, why
it is not zero and why it is on the order of the energy density
of matter) are solved in quantum liquids. The solution does not
depend on the details of the structure of the quantum liquid and
follows from the thermodynamic relation -- the Gibbs-Duhem
relation -- applied to the quantum vacuum as a medium.

\section{Black hole on brane}

\subsection{Simulation of Painlev\'e-Gullstrand metric}

Let us start with the modification of the Unruh dumb hole
\cite{Unruh} to superfluids. The effective metric experienced by
bosonic or fermionic quasiparticles moving in the background of
the superfluid vacuum which flows with the velocity
${\bf v}({\bf r})$ is
\begin{equation}
  ds^2=-dt^2 + {1\over c^2} \left(d{\bf r}-{\bf v}({\bf
r})dt\right)^2 +{1\over c^2} r^2d\Omega^2~.
\label{Painleve}
\end{equation}
Here $c$ is the maximum attainable speed of the low-energy
quasiparticle. If one considers the radial flow of the superfluid
vacuum with the following velocity field
\begin{equation}
{\bf v}({\bf r})=\pm \hat{\bf r} c\sqrt{r_h\over r}
~,~r_h={2MG\over c^2}~,
\label{VelocityField}
\end{equation}
one obtains the metric corresponding to the Painlev\'e-Gullstrand
spacetime describing the gravitational field
produced by the point source of mass $M$.  At $r=r_h$ where the
flow velocity reaches $c$ one has the black hole horizon in the
case of the vacuum flowing inward (sign -) and the white hole
horizon in the case of the flow outward (sign +)
\cite{Unruh,Visser}). In superfluids, a flow of the
superfluid vacuum violates the time-reversal symmetry: the time
reversal operation reverses the direction of flow, $T{\bf
v}=-{\bf v}$, and thus transforms the black hole into the white
one.

  The
Painlev\'e-Gullstrand type metric describes the space-time both
in exterior and interior regions, and this spacetime, though not
static, is stationary. That is why it has many advantages (see
  \cite{Martel,ParikhWilczek,Doran}), and in
particular these properties of the Painlev\'e-Gullstrand
metric  allow us to discuss the behavior of the quantum vacuum
in the interior of the black hole.

For many reason the radial flow with horizon cannot be realized
in normal liquids and also in superfluids, as well as many other
flow field configurations suggested for the condensed
matter simulation of horizons. Here we discuss the most
realistic situation, when the pair of white-hole and black-hole
horizons can be formed in quantum liquids and they can live for
reasonably long time.
This scenario of formation of horizons is based on recent
experiments with two superfluids sliding along each other
\cite{Kelvin--HelmholtzInstabilitySuperfluids,Brane}.

\subsection{Simulation of horizons and physical singularities on
brane}

In an applied gradient of magnetic field two superfluids,
$^3$He-A and $^3$He-B, are separated by the AB-brane. They can
move without any dissipation along the brane  with different
velocities
${\bf v}_1$ and ${\bf v}_2$. Thes velocities are determined in
the container frame where the flow field is stationary. From the
hydrodynamic equations it follows that that the quasiparticles
living on AB-brane -- the surface waves called ripplons -- have
the following dispersion relation if the container has the form
of a thin slab
\begin{eqnarray}
&\alpha_1  (\omega- {\bf p}\cdot {\bf v}_1)^2
+ \alpha_2(\omega- {\bf p}\cdot {\bf v}_2)^2=&\nonumber\\
&= c^2p^2\left(1+ {p^2\over p_0^2}\right) -   2i
\Gamma(p)
\omega,&
\label{ThinSlabsSpectrum}
\end{eqnarray}
where
\begin{eqnarray}
\alpha_1  = \frac{h_2\rho_1}{ h_2\rho_1+h_1\rho_2}~,~
\alpha_2  = \frac{h_1\rho_2}{ h_2\rho_1+h_1\rho_2}~,~\alpha_1
+\alpha_2=1~,
\label{ThinSlabsSpectrumParameters2}
\end{eqnarray}
\begin{equation}
p_0^2= {F\over\sigma},~~~c^2=
   F{h_1 h_2\over \rho_1 h_2+\rho_2 h_1}~.
\label{ThinSlabsSpectrumParameters2}
\end{equation}

Here $\sigma$ is the surface tension of the AB-brane; $F$ is the
force stabilizing the position of the brane -- an applied
magnetic-field gradient (see Ref.
\cite{Kelvin--HelmholtzInstabilitySuperfluids}); $\rho_1$ and
$\rho_2$ are mass densities of the liquids; $h_1$ and $h_2$ are
thiknesses of the layeres of A and B phases, and we assume the
`shallow water' condition
$ph_1\ll 1$ and
$ph_2\ll 1$; finally $\Gamma(p)$ is the coefficient in front of
the friction force experienced
by the AB-brane when it moves with respect
to the 3D environment along the normal. This friction term which
couples the 2D  brane with the 3D environment violates  Galilean
invariance in the 2D world of the AB-brane. In the low-energy
effective theory this term is responsible for the violation of
the Lorentz invariance in the brane world caused by the
interaction with the higher dimensional environment.

For $p\ll p_0$ the main part of Eq.
(\ref{ThinSlabsSpectrum}) can be
rewritten in the Lorentzian form
\begin{eqnarray}
g^{\mu\nu}p_\mu p_\nu= 2i\omega\Gamma(p) -c^2p^4/p_0^2,
\label{PGwithDissipationAndNonlinear}
\\
p_\mu=(-\omega,p_x,p_y),~~~p =\sqrt{p_x^2+p_y^2},
\label{PGwithDissipationAndNonlinear2}
\end{eqnarray}
while the right-hand side of Eq.(\ref{PGwithDissipationAndNonlinear})
contains the remaining small terms violating Lorentz invariance  --
attenuation of ripplons due to the friction and their nonlinear
dispersion. Both terms come from the physics which is `trans-
Planckian'
for the ripplons.  The quantities
$p_0$ and
$cp_0$ play the role of the Planck momentum and Planck
energy within the brane: they determine the scales where the
Lorentz symmetry is
violated. The  Planck scales of the 2D physics in brane are actually
much
smaller than the `Planck momentum' and `Planck energy' in the 3D
superfluids outside the brane. The function $\Gamma(p)$ is
determined by the physics of quasiparticles living in
bulk which scatter on the brane.

At sufficiently small $p$ both non-Lorentzian terms~--
attenuation and nonlinear dispersion on the right-hand side of
Eq.(\ref{PGwithDissipationAndNonlinear})~-- can be ignored, and the
dynamics of ripplons living on the AB-brane is described by
the following effective contravariant metric
$g^{\mu\nu}$:
\begin{equation}
\begin{array}{c}
g^{00}=-1,~~~g^{0i}=-\alpha_1 v_1^i - \alpha_2
v_2^i,\\[1mm]
g^{ij}=c^2\delta^{ij}-\alpha_1 v_1^iv_1^j - \alpha_2 v_2^i
v_2^j.\end{array}
\label{ThinSlabsRipplonContravMetric}
\end{equation}
Introducing relative velocity ${\bf U}$ and the  mean velocity
${\bf v}$
of two superfluids:
\begin{equation}
   {\bf v}=\alpha_1{\bf v}_1 +\alpha_2{\bf
v}_2,~ ~{\bf U}=  \sqrt{\alpha_1\alpha_2}\left( {\bf v}_1 - {\bf
v}_2\right),
\label{NewVelocities}
\end{equation}
one obtains the following expression for the effective contravariant
metric
\begin{equation}
g^{00}=-1~,~~
g^{0i}=-v^i~~,~~g^{ij}=c^2\delta^{ij}-v^iv^j -U^i U^j~.
\label{ThinSlabsRipplonContravMetric2}
\end{equation}
The interval of the effective 2+1 spacetime where  ripplons move
along geodesic curves is
\begin{eqnarray}
\nonumber ds^2=g_{\mu\nu}dx^\mu dx^\nu=\\
\nonumber -~dt^2\left(1-{v^2\over c^2}-{
({\bf U}\cdot {\bf v})^2\over c^2(c^2-U^2)} \right)-~2dt
\left( {{\bf v}\cdot d{\bf r} \over c^2} + {({\bf U}\cdot d{\bf
r})({\bf U}\cdot {\bf v})
\over c^2(c^2-U^2)} \right)\\
  +{ d{\bf r}^2\over c^2} + {({\bf U}\cdot
d{\bf r})^2
\over c^2(c^2-U^2)}  ~.
\label{Interval}
\end{eqnarray}
If $U=0$, the two superfluids move with the same
velocity and thus can be represented as a single liquid, as a
result the Eq.(\ref{Painleve}) is restored. But in the general
case (\ref{Interval}) the system is richer than in the case of a
single liquid. In principle all the interesting 2+1 metrics can
be constructed using two velocity fields. In particular, the
ergosurface (which is the ergoline in the 2+1 system) is
obtained when
\begin{equation}
  g_{00}=0~~,~~{\rm or}~~{v^2\over c^2} +{ ({\bf U}\cdot {\bf
v})^2\over c^2(c^2-U^2)}=1~,
\label{Ergosurface}
\end{equation}
and the physical singularity is obtained at $U=c$ where the
determinant of the metric
\begin{equation}
g=-{1\over c^2(c^2-U^2)}
\label{CovarianAcousticMetricReducedAnisotropic}
\end{equation}
changes sign.
In the particular case when ${\bf U}\perp {\bf v}$ and $v<U$, the
singularity at
$U=c$ is  naked  violating the `cosmic censorship' in
quantum liquids.

The arrangement of the planned experiment will be similar to that
suggested by Sch\"utzhold and Unruh \cite{SchutzholdUnruh}.
The A and B vacua with the AB-brane between them are confined
in the channel. The shape of the cross-section of the channel is
such that in the rotating cryostat the superflluid
velocities are inhomeheneous as is shown in the Figure 1.
In the typical situation the A-phase is at rest in the container
frame, while the B-phase is at rest in the inertial frame, which
means that ${\bf U}= {\bf v}=\hat {\bf \phi}v(\phi)$ and the
interval of the effective 2+1 spacetime
in which ripplons move along the geodesic curves is given by
\begin{eqnarray}
ds^2=
-d\tilde t^2 {c^2 - 2v^2(\phi)\over
c^2 -
v^2(\phi)} +{r^2d\phi^2 \over c^2 - 2v^2(\phi)} +dr^2~,
\label{Interval1}
\\
  d\tilde t =dt + {v(\phi)rd\phi\over c^2 -
2v^2(\phi)}  ~.
\label{Interval3}
\end{eqnarray}
In this geometry there are two horizons, the black-hole and
white-hole,  at points where $v^2_{\rm hor}=c^2/2$ and thus
$g_{00}=0$. The physical singularities occur at points where
$v^2_{\rm singularity}=c^2$ and thus
$g_{00}=\infty$ (see Figure 1).

\subsection{Vacuum instability due to interaction with bulk}

\begin{figure}
\centerline{\includegraphics[width=0.8\linewidth]{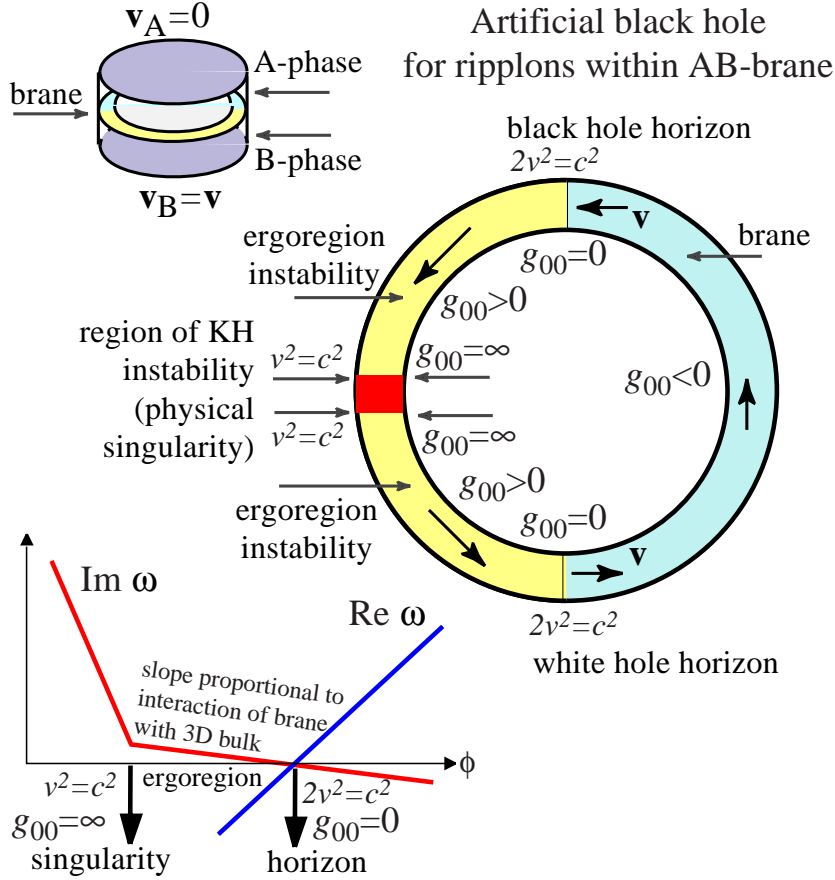}}
   \caption{The AB-brane -- the ring-shaped interface between two
bulk quantum vacua. The coordinate singularities in the
effective metric for ripplons living on brane occur at two lines
where   $v^2=c^2/2$ and thus $g_{00}=0$. These two lines
represent the black-hole and white-hole horizons. The physical
singularities occur at two lines where $v^2=c^2$ and thus
$g_{00}=\infty$.
  {\it Bottom left}: real and imaginary parts of the
ripplon spectrum. At the horizons both of them change sign. The
attenuation of ripplons outside the horizons transforms to
the amplification beyond the horizon, which means that the
quantum vacuum becomes unstable in the interior of the black
hole, and as a result the interior region shrinks. If the
interaction with the bulk environment is weak, the vacuum beyond
the horizon is long lived, and the physical singularity can be
also reached and investigated. }
   \label{HorizonABbraneTangFig}
\end{figure}

The important feature of the brane physics is the interaction of
the brane matter (ripplons) with the bulk world. In the
superfluid
$^3$He this interaction gives rise to the imaginary part of the
spectrum of brane quasiparticles in
Eq.(\ref{PGwithDissipationAndNonlinear}): the energy of ripplons
has a finite dissipation rate. In the presence of a horizon the
situation changes drastically: When the horizon is crossed and
the real part of the quasiparticle energy changes sign according
to Eq.(\ref{PGwithDissipationAndNonlinear}), the imaginary part
of the spectrum also changes sign (see Figure 1 {\it bottom
left}). This means that the attenuation of perturbations which
occurs outside the horizon transforms to the amplification of
perturbations beyond the horizon, i.e. due to the interaction
with the bulk environment the quantum vacuum of the brane
becomes unstable in the interior region of the black hole. This
instability finally leads to the shrinking of the interior
region.

In superfluid $^3$He the rate of the brane vacuum decay is
determined both by the brane and bulk parameters:
\begin{equation}
{1\over \tau} \sim \Gamma(p) \sim p^2 {c_{\rm bulk}T^3\over
p_{\rm brane} E_{\rm bulk}^3}~,
\label{VacuumDecay}
\end{equation}
where $T$ is the temperature in the bulk; $c_{\rm bulk}$ is the
maximum attainable speed for quasiparticles living in bulk;
$E_{\rm bulk}$ is the Planck energy scale in bulk; and $p_{\rm
brane}=p_0$ is the brane Planck momentum. Typically
the bulk parameters are much larger than that on the brane:
$c_{\rm bulk}\gg c_{\rm brane}$; $p_{\rm
bulk}\gg p_{\rm brane}$; and $E_{\rm
bulk}\gg E_{\rm brane}\equiv c_{\rm brane}p_{\rm brane}=cp_0$.

In the current experiments with the AB-brane
\cite{Kelvin--HelmholtzInstabilitySuperfluids},
the shallow water conditions are not satisfied, and thus the
ripplons are non-relativistic and are not described by the
effective metric. That is why the horizon was not simulated
in these experiments. However, the notion of the ergoregion --
the region where the ripplons have negative energy in the
container frame -- is well defined even for the non-relativistic
case. The Figure 1 {\it bottom
left} properly describes the spectrum of the
non-relativistic ripplons too, if one substitutes the horizons by
the ergosurfaces. The physical singularity corresponds to the
points where the Kelvin-Helmholtz instability takes place.
The vacuum decay in the ergoregion occurs
essentially in the same way as beyond the horizon, and the
experimental data for the threshold of the ergoregion
instability are in an excelent agreement with the theory. Thus
the non-relativistic analog of the discussed mechanism of the
brane vacuum decay due to the interaction with the
higher-dimensional environment has been experimentally confirmed.

Something similar can occur for the astronomical black holes if
we live on the brane of the multidimensional world. The
interaction of our brane matter with the bulk environment can
lead to the decay of the quantum vacuum beyond the black hole
horizon. The time of
$\tau$ of the decay depends on the ratio of the bulk and brane
parameters, and in principle, this time can be shorter that the
quantum process of the black hole evaporation introduced by
Hawking. The goal of the future experiments with the AB-brane is
to study the vacuum decay in
the relativistic regime when the horizon is developed, and at low
temperature where the decay time is long, so that the
other processes of the vacuum decay become relevant, such as
the Corley--Jacobson lasing scenario
\cite{BHlaser} which occurs in the presence  of the pair of
horizons.

\section{Entropy of extremal black hole}

\subsection{Simulation of extremal and nearly extremal black
holes}

In previous Section it was shown that the event horizon for the
relativistic quasiparticles can be constructed in quantum
liquids at least in principle. Since the microstates of the
quantum liquids are well known we can study the problem of the
thermodynamics in the presence of the horizon.

\begin{figure}
\centerline{\includegraphics[width=0.9\linewidth]{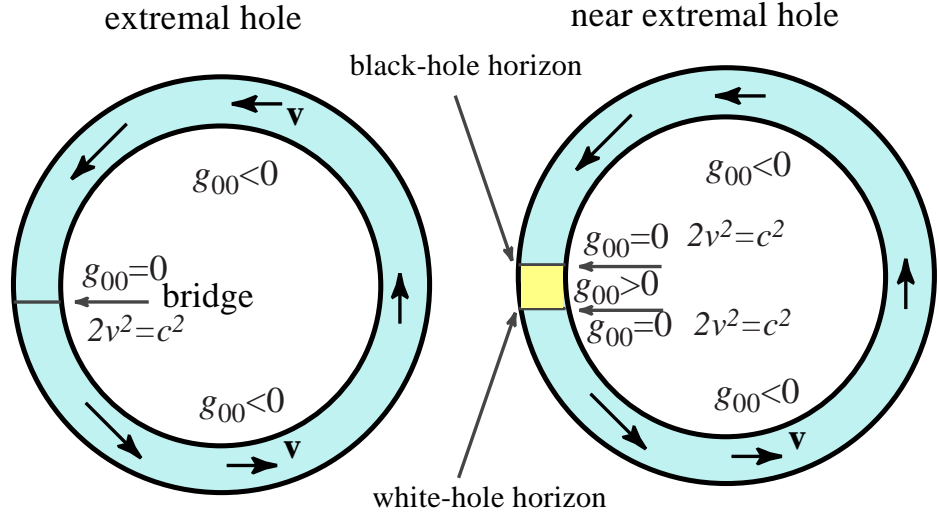}}
   \caption{{\it
Right}: When the black and white horizons in Fig.
\protect\ref{HorizonABbraneTangFig} are close to each other the
metric becomes similar to that of the nearly extremal hole. {\it
Left}: When the two horizons merge the effective metric
corresponds to that of the extremal black hole.}
   \label{ExtremalHoleFig}
\end{figure}

When the two horizons are close to each other (Figure 2 {\it
right}) the quasiparticle interval in Eq.(\ref{Interval1})
becomes
\begin{equation}
ds^2=
-d\tilde t^2 (\phi^2-\phi_{\rm hor}^2) +{r^2d\phi^2 \over
\phi^2-\phi_{\rm hor}^2} +dr^2~.
\label{Interval4}
\end{equation}
When the two horizons merge  (Figure 2 {\it
left}) one has $\phi_{\rm hor}^2=0$, and the interval becomes
\begin{equation}
ds^2=
-d\tilde t^2 \phi^2  +{r^2d\phi^2 \over
\phi^2} +dr^2~,
\label{Interval5}
\end{equation}
One can compare this with the interval for matter in the
presence of the Reissner-Nordstr\"om black hole of mass
$M$ and electric charge $Q$
\begin{eqnarray}
ds^2=
-d\tilde t^2{(r-r_+)(r-r_-)\over r^2}+
dr^2{r^2\over(r-r_+)(r-r_-)}+ r^2d\Omega~,
\label{ReissnerPainleveForm1}
\\ r_+r_-=Q^2~,~r_++r_-=2M G~,
\label{ReissnerPainleveForm2}
\end{eqnarray}
where $r_+$ and
$r_-$ are external and inner horizons. For the nearly
extremal black hole, the two horizons are close to each other,
$r_+-r_-\ll r_++r_-\equiv 2r_h$. Introducing the
coordinate $\tilde r =r- (r_++r_-)/2$ in the vicinity of
horizons and the quantity
$\tilde r_h^2=(r_+-r_-)^2/4$ one obtains
\begin{equation}
ds^2\approx
-d\tilde t^2{\tilde r^2 -\tilde r_h^2\over r_h^2}+
d\tilde r^2{r_h^2\over\tilde r^2 -\tilde r_h^2}+ r_h^2d\Omega~,
\label{NearExtr}
\end{equation}
Thus close to the horizons the
radial part of the Reissner-Nordstr\"om metric is similar to the
azimuthal part of the quasiparticle metric in Eq.
(\ref{Interval4}). When the two horizons merge, i.e. when
$\tilde r_h=0$, one obtains the metric of the extremal hole
\begin{equation}
ds^2\approx
-d\tilde t^2{\tilde r^2\over r_h^2}+
d\tilde r^2{r_h^2\over\tilde r^2}+ r_h^2d\Omega~,
\label{Extr}
\end{equation}
which is thus simulated by the quasiparticle interval in
Eq.(\ref{Interval5}). Since the thermodynamics of the nearly
extremal and the extremal holes are determined by the
microstates in the vicinity of the horizons, the difference
between the metrics in Eqs.(\ref{NearExtr}--\ref{Extr}) and
their quasiparticle counterparts in
Eqs.(\ref{Interval4}--\ref{Interval5}) is not important.

The above examples demonstrate that in principle it is possible
to simulate  different types of horizons in quantum liquids.
Let us now assume that we can construct in
a quantum liquid the following effective metric for
the `relativistic' quasiparticles:
\begin{eqnarray}
  ds^2
={1\over b(r)}\left[- c^2(r)dt^2 + \left(dr -
v(r)dt\right)^2\right] +r^2d\Omega^2 =
\label{PainleveModified2}\\
={1\over b(r)}\left[-(c^2(r) - v^2(r))  d\tilde t^2
+ dr^2{c^2(r)\over c^2(r) - v^2(r)}\right]+r^2d\Omega^2~,
\label{PainleveModified3}
\\
  d\tilde t =dt + {v(r)dr\over c^2(r)-v^2(r)}    ~.
\label{PainleveModified4}
\end{eqnarray}
Here $v(r)$ is the velocity of the radial flow of the liquid;
$c(r)$ is the local maximum attainable speed of the relativistic
quasiparticle in the reference frame comoving with the liquid;
and $b(r)$ is some function related to the local  density of the
liquid. One can check that the metric in
Eq.(\ref{PainleveModified3}) coincides with
the metric of Reissner-Nordstr\"om black hole of mass
$M$ and electric charge $Q$ in Eq.(\ref{ReissnerPainleveForm1})
if one chooses
\begin{equation}
b(r)={c(r)\over c}=1+{Q^2\over r^2}~~,~~v(r)=-
\sqrt{2GMb(r)\over r}~~,~~c= c(r=\infty)~.
\label{VelocityFieldModified}
\end{equation}

The quasiparticle metric in Eq.(\ref{PainleveModified2}) is thus
the analog of the Painlev\'e-Gullstrand metric describing the
spacetime both in exterior and interior regions of the
Reissner-Nordstr\"om black hole.

\subsection{Fermi surface in the interior of nearly extremal
black hole}

The spacetime in Eq.(\ref{PainleveModified2}), though not
static, is stationary. That is why the quasiparticle energy
spectrum in the interior region is well determined.
It has the form of Eq.(\ref{PGwithDissipationAndNonlinear})
\begin{equation}
g^{\mu\nu}p_\mu p_\nu+m^2+c^2p^4/p_0^2=0~,
\label{PGwithDissipationAndNonlinear5}
\end{equation}
where the effective Painlev\'e-Gullstrand metric $g^{\mu\nu}$ is
given by Eq.(\ref{PainleveModified2}), and we neglected the
possible interaction of quasiparticles with the environment which
leads to the broadening of the spectrum. The dispersion relation
(\ref{PGwithDissipationAndNonlinear5}) contains only the first
non-linear correction to the relativistic spectrum which
violates the Lorentz symmetry. Though the main contribution to
the thermodynamics of the black hole comes from the short wave
length, where the Planck physics intervenes, it happens that for
the nearly extremal and extremal holes  only this first
correction is important, while the higher-order terms can be
neglected. Thus one has the following quasiclassical
energy spectrum $E({\bf p})$ in the background of
the Reissner-Nordstr\"om black hole:
\begin{equation}
  {1\over b(r)}\left(E-p_r v(r)\right)^2 = b(r)
c^2p_r^2 +c^2p_\perp^2 +m^2+ {p^4\over p_0^2}~.
\label{SuperluminalSpectrumModified}
\end{equation}
where   $p_r$ is the radial
momentum of the (quasi)particle and $p_\perp$ is its transverse
momentum.

We are interested in the  spectrum  of fermionic quasiparticles,
since according to the quantum liquid analogy for the quantum
vacuum only the fermions are fundamental while the bosons
appear as collective modes of the quantum vacuum
\cite{VolovikBook}. In the Standard Model these quasiparticles
correspond to
$N_F=16N_g$ chiral fermionic species where
$N_g$ is the number of generations. The most important property
of the fermionic spectrum in the presence of the horizons is the
existence of the Fermi surface, which appears even for fermions
with nonzero mass
$m$: the Fermi surface appears at large momentum $p$ where the
mass term can be neglected. Fermi surface is the 2D surface in
the 3D momentum space, where the energy of the particle is zero:
$E({\bf p})=0$. For the spectrum in
Eq.(\ref{SuperluminalSpectrumModified}) this 2D manifold of
fermion zero modes is given by equation, which expresses the
radial momentum in terms of the transverse momentum $p_\perp$
(we put $c=1$):
\begin{equation}
   p_r^2(p_\perp)= {1\over 2}p_0^2
\left({v^2\over b}-b\right) -p_\perp^2
\pm \sqrt{{1\over 4}p_0^4 \left({v^2\over b}-b\right)^2
  +p_0^2 p_\perp^2\left(b-1-{v^2\over b}\right)}~.
\label{FermiSurface}
\end{equation}
  This surface exists at each point ${\bf r}$
between the horizons, where
$v^2>b^2$. It exists only in the restricted range of
the  transverse momenta, with the restriction provided by the
  parameter $p_0$ which plays the role of the Planck momentum:
\begin{equation}
    p_\perp^2<{1\over 4} p_0^2 { (v^2- b^2)^2 \over
b^2-b^3+bv^2} ~.
\label{MomentumRestriction}
\end{equation}
This means that the Fermi surface is a closed surface in the
3D momentum space ${\bf p}$.

\subsection{Entropy of nearly extremal black hole}

The main thermodynamic property of the fermion zero modes with
the Fermi surface is that its vacuum is characterized by a finite
density of fermionic states (DOS)  $N(E=0)$. For each of
$N_F=16N_g$ fermionic species the DOS is:
\begin{eqnarray}
N(E)= \sum_{{\bf p},{\bf r}} \delta ( E({\bf
p})- E)~,
\label{DOSdefinition1}\\
N( E=0)={4\pi
\over (2\pi)^3}
  \int_{r_-}^{r_+}r^2dr\int d^3p~
\delta ( E({\bf p}))=
  {  1\over  \pi }
  \int_{r_-}^{r_+}r^2dr\int_0^{p_\perp^2(r)}
{d(p_\perp^2)\over |v_G|}~,
\label{DOSdefinition3}
\end{eqnarray}
  where $v_G$ is the radial component of the group
velocity of (quasi)particles at the Fermi surface:
\begin{equation}
   v_G(E=0,p_\perp,r)={dE\over dp_r}\big|_{E=0}=\mp {1\over
v}\sqrt{(v^2-b^2)^2 +4{p_\perp^2\over
p_0^2}(b^3-b^2-bv^2) }~.
\label{GroupVelocity}
\end{equation}

Integration over $p_\perp^2$ in Eq.(\ref{DOSdefinition3})
gives for the DOS
\begin{equation}
N(E=0)={p_0^2\over \pi}\int_{r_-}^{r_+}r^2dr
{|v|(v^2-b^2)\over  bv^2+b^2-b^3}~.
\label{DOSClassical}
\end{equation}
In the extreme limit $Q\rightarrow M$ one obtains
\begin{equation}
N(E=0)\approx  {4 \over
3\pi}p_0^2r_+^3\left(1-{Q^2\over M^2}\right)^{3/2}~.
\label{DOSClassical4}
\end{equation}
Let us emphasize that the (quasi)particles contributing to this
DOS live between the two horizons, i.e. $r_-<r<r_+$.
The characterisic momenta of these (quasi)particles are
$p\sim p_0
\sqrt{1-{Q^2\over M^2}}$. In the extreme-hole limit these
momenta, though much larger than any fermionic mass, are still
much smaller than the Planck momentum $p_0$. This is the reason
why in this limit only the first non-linear correction to the
relativistic spectrum in
Eq.(\ref{PGwithDissipationAndNonlinear5}) is relevant.

At non-zero temperature the fermionic microstates forming the
Fermi surface give the following Bekenstein entropy:
\begin{equation}
{\cal S}(T,M,Q)
   ={\pi^2\over 3}  N(E=0)T= 4\pi N_F
{p_0^2r_+^3T\over \hbar^3 c} \left(1-{Q^2\over
M^2}\right)^{3/2}  ~,
\label{ThermalEntropy}
\end{equation}
which is linear in $T$.

\subsection{Entropy of extremal black hole}

For the extremal black hole ($Q=M$) the two horizons merge and
the Fermi surface disappears. As a result the linear in $T$
term  in Eq.(\ref{ThermalEntropy}) for the Bekenstein
entropy becomes zero. This means that the entropy of the
extremal black hole must be of higher order in $T$. Let us
calculate this entropy. It can be obtained directly from the
thermal energy
\begin{equation}
{\cal E}=\int  dE E N(E)f(E/T)=\sum_{{\bf p},{\bf
r}}E({\bf p},{\bf r})f(E({\bf p},{\bf r})/T)~,
\end{equation}
where $f$ is the Fermi distribution function.
  For the extremal hole one has
$b^2(r)-  v^2(r)\approx 2(r-r_+)^2/r_+^2$, and after the
change of variables
\begin{equation}
    {p_r(r-r_+)^2\over 2r_+^2}=x^2~~,~~
    {p_\perp^2\over 2p_r}=y^2~~,~~    {p^3_r\over
2p_0^2}=z^2~,~E=x^2+y^2+z^2=u^2~,
\label{ChangeVariables}
\end{equation}
  one obtains the following dependence of the DOS on $E$
\begin{equation}
  N(E)dE={d^3rd^3p\over (2\pi)^3}={2\over \pi}dp_r~ d(p_\perp^2)
~r^2dr ={1\over 3\pi}p_0 r_+^3 |u|d^3u={2\over 3c^2}p_0
r_+^3 E dE~.
\label{VolumeEelement}
\end{equation}
Thus the DOS of fermion zero modes living at the extremal black
hole is equivalent to the DOS of the 2D relativistic gas
of massless fermions with zero chemical potential, whose density
of states is
$N(E)=(2/3)p_0 r_+^3 E$.  The entropy of such gas and thus the
entropy of extremal black hole is
\begin{equation}
{\cal S}_{\rm extremal~hole}=3\zeta(3) N_F{p_0r_+^3\over
c^2\hbar^3} T^2~.
\label{EntropyExtremal}
\end{equation}
The entropy in Eq.(\ref{ThermalEntropy}) for the nearly extremal
black hole which is valid for
\begin{equation}
1\gg  1-{Q^2\over M^2} \gg \left({T\over cp_0}\right)^{2/3}  ~,
\label{Limits}
\end{equation}
smoothly transforms to the entropy  in
Eq.(\ref{EntropyExtremal}) for the extremal black hole, when the
parameter $1-{Q^2\over M^2}$ approaches $(T/cp_0)^{2/3}$.

Let us stress again that for the extremal black hole the main
contribution to DOS and thus to the thermodynamics comes from
the momenta  much smaller than the cut-off momentum $p_0$:
\begin{equation}
{p_r\over p_0} \sim \left({T\over cp_0}\right)^{1/3}\ll
1~~,~~{p_\perp\over p_0} \sim \left({T\over cp_0}\right)^{2/3}\ll
1~.
\label{MomentumIsSmall}
\end{equation}
That is why we do not need to know the trans-Planckian
physics, since only the first nonlinear correction to the
relativistic particle spectrum is important. The region in the
vicinity of the horizon which gives the main contribution to
the thermodynamics has the size
\begin{equation}
\Delta r=|r-r_+| \sim r_+\left({T\over cp_0}\right)^{1/3} ~.
\label{RegionIsSmall}
\end{equation}
For all reasonable temperatures the thickness of this shell is
much bigger than the Planck length $\hbar/p_0$. The
quasiclassical approximation which we used is valid when
$p_r\Delta r\gg 1$, i.e. when the temperature $T\gg
(cp_0)^{5/2}M^{-3/2}$, where $M$ is the mass of the black hole.
The quantum limit, when the quantum properties of the black hole
become important, is approached at temperature
$T_{\rm quantum} =(cp_0)^{5/2}M^{-3/2}$. This $T_{\rm quantum}$
is by factor
$(cp_0/M)^{1/2}$ smaller than the characteristic Hawking scale
$T_{\rm H} \sim  (cp_0)^2/M$, which demonstrates that the
non-linear dispersion of the energy spectrum drastically changes
the black-hole thermodynamics.

\section{Thermodynamics of the quantum vacuum}

Recent cosmological observations indicate that the cosmological
constant in the present Universe is non-zero and its magnitude is
by 123 orders of magnitude smaller than its natural scale
dictated by the Planck energy cut-off (see the recent review
paper \cite{Padmanabhan})
\begin{equation}
\Lambda_{\rm
nat}\sim {cp_0^4\over \hbar^3}~.
\label{LambdaNatural}
\end{equation}
  Such a huge disparity between
the theoretically expected and experimentally observed values
suggests that there must be some fundamental principle which
requires
$\Lambda$ to vanish almost completely. The quantum-liquid
analogs of the quantum vacuum give some hint how this almost
complete cancellation can occur without any fine tuning.  In
quantum liquids such a cancellation does occur, it follows from
the thermodynamics of the quantum liquid and does not depend on
the details of the microscopic (trans-Planckian) physics. That is
why one can expect that considering the thermodynamics of
the vacuum as an effective medium one obtains the same
cancellation.

Let us consider such quantum liquids, whose
low-frequency dynamics of collective modes is similar to that of
the quantum fields of our vacuum. An example is provided by
the superfluid $^3$He-A whose fermionic and  bosonic
quasiparticles have the same structure as chiral fermions, and
gauge and gravitational fields. The reason for such an analogy is
the momentum space topology which is common for the
ground state of $^3$He-A and for the quantum vacuum of the
Standard Model \cite{VolovikBook}.

Let us suppose that the quantum liquid is perfect in a
sense that in the low-energy corner it fully reproduces the
Standard Model. Then from the point of view of the low-energy
observers living in this liquid the cosmological
constant must have the natural value in Eq.(\ref{LambdaNatural}).
Here $p_0$ is the corresponding Planck momentum cut-off, which
in $^3$He-A is on the order of Fermi momentum; $c$ is the
maximum attainable speed for the low-energy quasiparticles,
which changes from 3 cm/s to 100 m/s  depending on the direction
of propagation; and the sign $\pm$ depends on the fermionic and
bosonic content of the effective theory.

As distinct from our vacuum, in the quantum liquids we know not
only the effective theory, but also the  microscopic
(trans-Planckian) physics, and thus we can exactly calculate
the vacuum energy -- the energy of the ground state of
the liquid -- and thus to find the cosmological constant exactly.
If one calculates the energy density
$\epsilon$ of the liquid, one obtains the same estimate for
its magnitude as in Eq.(\ref{LambdaNatural}). But does the energy
$\epsilon$ of the liquid really correspond to the cosmological
constant? The further inspection indicates that the role of the
cosmological constant $\Lambda$ is played by the thermodynamic
potential $\tilde\epsilon= \epsilon -\mu n$, where $n=N/V$ is the
density of particles of the liquid and $\mu$ is their chemical
potential. One should not confuse particles which comprise
the vacuum (say, the $^3$He atoms) and quasiparticles which
are the excitations above the vacuum and which form the analog
of  matter in quantum liquids.

The reason for the identification of the cosmological constant
with the thermodynamic potential,
$\Lambda=\tilde\epsilon$, is that in the quantum liquids it is
just the gradient expansion of $\tilde\epsilon$ (not of
$\epsilon$) which gives rise to the effective Hamiltonian
${\cal H}_{\rm effective}$ for the effective gravity and
effective matter fields (bosonic and fermionic quasiparticles):
\begin{equation}
{\cal H}-\mu {\cal N}=V\tilde\epsilon   +  {\cal H}_{\rm
effective}~~,~~\tilde\epsilon={1\over V}\langle vac|{\cal H}-\mu
{\cal N}|vac\rangle ~.
\label{Expansion}
\end{equation}
Thus $\tilde\epsilon$ is the proper energy of the vacuum for
the effective theory. That this is the correct choice also
follows from the Gibbs-Duhem relation applied to the
equilibrium quantum liquid
\begin{equation}
P=TS+\mu n -\epsilon ~,
\label{GibbsDuhemRrelation}
\end{equation}
where $P$ is the pressure and $S$ is the entropy density. From
this relation one obtains that the vacuum, i.e. an
equilibrium liquid at $T=0$, has the following equation
of state
\begin{equation}
P_{\rm vac}=  -\tilde\epsilon_{\rm vac} ~.
\label{PressureEnergy}
\end{equation}
This is just the equation of state for the vacuum in general
relativity, which follows from the cosmological term in the
Einstein action. This confirms our identification of
the cosmological constant emerging in
the quantum liquid.

Let us find $\tilde\epsilon_{\rm vac}$ for the
liquid in its ground state in the situation when the liquid is
isolated from the environment, for example, for a droplet
of the liquid in space. Since in the absence of the interaction
with the environment the pressure of the liquid is zero, from
Eq.(\ref{PressureEnergy}) it follows that in complete
equilibrium and in the absence of any external perturbations
disturbing the ground state of the liquid, the energy
$\tilde\epsilon_{\rm vac}$ of the ground state is always zero.
Thus the thermodynamics of the quantum vacuum demonstrates that,
if the gravity is effective, the big energy of the vacuum (and
thus the big cosmological constant) is cancelled without any
fine tuning.

The next step is to demonstrate that in the perturbed vacuum the
relevant vacuum energy density $\tilde\epsilon_{\rm vac}$ is
slightly non-zero being proportional to the energy density of
perturbations. This can be seen on the simplest example of
non-zero temperature if it is small compared to the Planck
energy scale, $T\ll cp_0$ (this condition is probably
satisfied even during the Big Bang, which would mean that even
at the highest density of matter the quantum vacuum is still
close to its equilibrium state). At
$T\neq 0$ the thermal quasiparticles are excited, and they
represent the matter living in the quantum vacuum of the liquid.
 From the Gibbs-Duhem relation for matter (the system of thermal
quasiparticles)
\begin{equation}
  P_{\rm matter}= TS -{1\over V} E_{\rm matter} ~,
\label{GibbsDuhemRrelation}
\end{equation}
and from the condition that in an isolated liquid the total
pressure must be zero
\begin{equation}
  P_{\rm matter}+P_{\rm vac}=0 ~,
\label{TotalPressure}
\end{equation}
one obtains the response of the cosmological constant to the
thermally excited matter:
\begin{equation}
\Lambda=\tilde\epsilon_{\rm vac}=-P_{\rm vac}= P_{\rm matter}=TS
-{1\over V}E_{\rm matter} ~.
\label{ResponseToThermalMatter}
\end{equation}
Thus in this example the cosmological constant is on the order
of the free energy of matter, which is what the
cosmological observations suggest.

The other perturbations which disturb the zero value of the
cosmological constant are the space curvature, the
time-dependent processes such as an expansion of the Universe,
Casimir effect, etc. (see \cite{VolovikBook}). Thus from the
quantum-liquid analog of the quantum vacuum it follows that the
cosmological constant is not a constant but is the evolving
physical parameter, which responds to the combined action of
different perturbations of the vacuum state. Our goal is to find
the laws of its evolution, and probably the further exploitation
of the condensed-matter analogies will help us to solve this
cosmological `constant' problem.

\vfill\eject

\end{document}